\documentclass[twocolumn,epjc3]{svjour3}          % twocolumn
\RequirePackage[T1]{fontenc}
\usepackage{amssymb}
\usepackage{multirow}
\smartqed  % flush right qed marks, e.g. at end of proof

\RequirePackage{graphicx}
\RequirePackage{mathptmx}      % use Times fonts if available on your TeX system
\RequirePackage{flushend}
\RequirePackage[numbers,sort&compress]{natbib}
\RequirePackage[colorlinks,citecolor=blue,urlcolor=blue,linkcolor=blue]{hyperref}

\journalname{Eur. Phys. J. C}

\begin{document}
\title{New observational constraints on $f(T)$ cosmology from radio quasars}

\author{Jing-Zhao Qi \thanksref{addr1},
        \
        Shuo Cao \thanksref{e1,addr1},
        \
        Marek Biesiada \thanksref{addr1,addr2},
        \
        Xiaogang Zheng \thanksref{addr1,addr2}
        \
        and Zong-Hong Zhu \thanksref{addr1}
        }

\thankstext{e1}{corresponding author \\  e-mail: caoshuo@bnu.edu.cn}
\institute{Department of Astronomy, Beijing Normal University,
Beijing, 100875, China \label{addr1}
\and
Department of Astrophysics and Cosmology, Institute of Physics,
University of Silesia, Uniwersytecka 4, 40-007, Katowice, Poland \label{addr2}}

\date{Received: date / Accepted: date}

\maketitle

\begin{abstract}

Using a new recently compiled milliarcsecond compact radio data set
of 120 intermediate-luminosity quasars in the redshift range $0.46<
z <2.76$, whose statistical linear sizes show negligible dependence
on redshifts and intrinsic luminosity and thus represent standard
rulers in cosmology, we constrain three viable and most popular
$f(T)$ gravity models, where $T$ is the torsion scalar in
teleparallel gravity. Our analysis reveals that constraining power
of the quasars data (N=120) is comparable to the Union2.1 SN Ia data
(N=580) for all three $f(T)$ models. Together with other standard
ruler probes such as Cosmic Microwave Background and Baryon Acoustic
Oscillation distance measurements, the present value of the matter
density parameter $\Omega_m$ obtained by quasars is much lager than
that derived from other observations. For two of the models
considered ($f_1$CDM and $f_2$CDM) a small but noticeable deviation
from $\Lambda$CDM cosmology is present, while in the framework of
$f_3$CDM the effective equation of state may cross the phantom
divide line at lower redshifts. These results indicate that
intermediate-luminosity quasars could provide an effective
observational probe comparable to SN Ia at much higher redsifts, and
$f(T)$ gravity is a reasonable candidate for the modified gravity
theory.

\end{abstract}

\section{Introduction}

The current cosmic acceleration has been supported by many
independent astrophysical observations, including type Ia supernovae
(SN Ia) \cite{riess1998supernova}, large scale structure
\cite{tegmark2004cosmological}, cosmic microwave background (CMB)
anisotropy \cite{spergel2003wmap}, etc. A mysterious component with
negative pressure, dubbed as dark energy, has been proposed to
explain this phenomenon in the framework of Einstein's general
relativity, which gave birth to a large number of dark energy models
including the cosmological constant ($\Lambda$CDM), scalar field
theory \cite{Caldwell:2005tm,Zlatev:1998tr,Tsujikawa:2013fta}, and dynamical dark energy models
\cite{barboza2009generalized,maor2001limitations,linder2003exploring,Wei:2013jya,boisseau2000reconstruction}.
The other direction one could follow in search for solution of the
accelerating cosmic expansion enigma is to construct modified
theories of gravity instead of invoking exotic dark energy. Large
majority of works in this direction concentrated on the brane-world
Dvali-Gabadadze-Porrati (DGP) model \cite{dvali2000metastable},
$f(R)$ gravity \cite{chiba20031}, and Gauss-Bonnet gravity
\cite{nojiri2005modified}.

Equally well, one can also modify the gravity according to the
scenario described by the so-called $f(T)$ theory
\cite{bengochea2009dark}, which was proposedin the framework of the
Teleparallel Equivalent of General Relativity (also known as
Teleparallel Gravity). In this approach, the Levi-Civita connection
used in Einstein's general relativity is replaced by the
Weitzenb\"ock connection with torsion, while the Lagrangian density
of this theory is the torsion scalar $T$. Compared with the $f(R)$
theory leading to the fourth order equations, the field equations of
the $f(T)$ theory are in the form of second order differential
equations, which provides an important advantage of this approach.
In addition, if certain conditions are satisfied, the behavior of
$f(T)$ cosmologies is similar to several popular dark energy models,
such as quintessence \cite{xu2012phase}, phantom \cite{wu2011f}, DGP
model \cite{cai2015f} and transient acceleration
\cite{qi2016transient}. Due to the above mentioned property, $f(T)$
theory and its cosmological applications has gained a lot interest
in the literature. The
detailed introduction to the $f(T)$ theory could be found in %Ref.
\cite{cai2015f,yang2011new}.

In this paper, we focus on using the currently released quasar data
\cite{cao2016measuring} to provide the constraints on various $f(T)$
gravity models. Recently, the angular size of compact structure in
radio quasars versus redshift data from the very-long-baseline
interferometry (VLBI) observations have become an effective probe in
cosmology. Reliable standard rulers and standard candles at
cosmological scales are crucial for measuring cosmic distances at
different redshifts. For instance, the type Ia supernovae are
regarded as standard candles, while the BAO peak location is
commonly recognized as a fixed comoving ruler. The increasing
observational material concerning these two distance indicators has
been widely used in various cosmological studies. In the past, there
were controversial discussions about whether the compact radio
sources could act as standard rulers
\cite{jackson1997deceleration,vishwakarma2001consequences,
lima2002dark,zhu2002cardassian,chen2003cosmological}. The difficulty
lies in the fact that the linear sizes $l_m$ of compact radio
sources might not be constant, i.e., its value is dependent on both
redshifts and some intrinsic properties of the source (luminosity,
for example). Based on a 2.29 GHz VLBI all-sky survey of 613
milliarcsecond ultra-compact radio sources
\cite{kellermann1993cosmological,gurvits1994apparent}, Cao et al.
\cite{cao2016measuring} presented a method to divide the full sample
into different sub-samples, according to their optical counterparts
and luminosity (low-luminosity quasars, intermediate-luminosity
quasars, and high-luminosity quasars). The final results indicated
that intermediate-luminosity quasars show negligible dependence on
both redshifts $z$ and intrinsic luminosity $L$, which makes them a
fixed comoving-length standard ruler.
% in the standard model of cosmology.
More recently, based on a cosmological-model-independent method to
calibrate the linear sizes $l_m$ of intermediate-luminosity quasars,
Cao et al. \cite{cao2016cosmology} investigated the cosmological
application of this data set and obtained stringent constraints on
both the matter density $\Omega_m$ and the Hubble constant $H_0$,
which agree very with the recent \textit{Planck} results. The
advantage of this data set, compared with other standard rulers: BAO
\cite{percival2010baryon,blake2011wigglez,beutler20116df}, clusters
\cite{bonamente2006determination}, strong lensing systems
\cite{biesiada2011dark,cao2012SL,cao2015SL}), is that quasars are
observed at much higher redshifts ($z\sim 3.0$). Therefore, it may
be rewarding to test the $f(T)$ theory with this newly revised
quasar data. In this paper, we examine constraints on the viable
$f(T)$ cosmological models imposed by the quasars. We compare them
with analogous results obtained with the newly revised Union2.1 set
--- the largest published and spectroscopically confirmed SN Ia
sample to date. We expect that different systematics and
sensitivities of these two different probes (rulers vs. candles) can
give complementary results on the $f(T)$ theory.

This paper is organized as follows: In section 2 we briefly
introduce the $f(T)$ gravity and its cosmological consequences. In
section 3 we present the latest data sets for our analysis and
perform a Markov chain Monte Carlo analysis using different data
sets. Finally, we summarize the main conclusions in Section 4.

\section{The $f(T)$ theory}
\label{fT theory}

In this section we brief review the $f(T)$ gravity in the framework
of cosmology, and then present three specific $f(T)$ models to be
analyzed in this work.

\subsection{The $f(T)$ cosmology}
We use the vierbein fields ${\mathbf{e}_i(x^\mu)}$ ($i=0, 1, 2, 3$),
which is an orthonormal basis for the tangent space at each point
$x^\mu$ of the manifold $\mathbf{e} _i\cdot\mathbf{e}_j=\eta_{i\,
j}$, and whose components are $e^\mu_i$ $(\mu=0,1,2,3)$ (here Latin
indices stand for the tangent space and Greek indices refer to the
manifold). Its dual vierbein gives the metric tensor
$g_{\mu\nu}(x)=\eta_{i\, j}\, e^i_\mu (x)\, e^j_\nu (x)$. In $f(T)$
theory, instead of the torsionless Levi-Civita connection in
Einstein's General Relativity, the curvatureless Weitzenb\"ock
connection is considered, and hence the torsion tensor describing
the gravitational field is
\begin{equation}
\label{torsion2} T^\lambda_{\mu\nu}\equiv e^\lambda_i(\partial_\mu
e^i_\nu-\partial_\nu e^i_\mu).
\end{equation}

The Lagrangian of teleparallel gravity is constructed by the torsion
scalar as \cite{bengochea2009dark}
\begin{equation}  \label{lagTele}
T\equiv S_\rho^{\mu\nu}\:T^\rho_{\mu\nu},
\end{equation}
where
\begin{equation}  \label{S}
S_\rho^{\mu\nu}=\frac{1}{2}\Big(K^{\mu\nu}_{\rho}+\delta^\mu_%
\rho T^{\theta\nu}_{\theta}-\delta^\nu_\rho
T^{\theta\mu}_{\theta}\Big),
\end{equation}
and the contorsion tensor $K^{\mu\nu}_{\rho}$ is given by
\begin{equation}  \label{K}
K^{\mu\nu}_{\rho}=-\frac{1}{2}\Big(T^{\mu\nu}_{\rho}
-T^{\nu\mu}_{\rho}-T_{\rho}^{\mu\nu}\Big).
\end{equation}
In the $f(T)$ theory, the Lagrangian density is a function of $T$
\cite{bengochea2009dark}, and the action reads
\begin{equation}  \label{accionTP}
I= \frac{1}{16\, \pi\, G}\, \int d^4x\:e\:f(T),
\end{equation}
where $e=det(e^i_\mu)=\sqrt{-g}$. The corresponding field equation
is
\begin{eqnarray}\nonumber
[e^{-1}\partial_\mu(eS_i^{\mu\nu})-e_i^{\lambda}
\:T^\rho_{\mu\lambda}S_\rho^{\nu\mu}]f_T &+& \\
 S_i^{\mu\nu}\partial_\mu T f_{TT}
+\frac{1}{4}e_i^\nu f(T)&=&\frac{1}{2}k^2e_i^{\rho}T_\rho^{\nu},
\label{ecsmovim}
\end{eqnarray}
where $k^2=8\pi G$, $f_T\equiv df/dT$, $f_{TT} \equiv d^2f/dT^2$,
$S_i^{\mu\nu}\equiv e_i^{\rho}S_\rho^{\mu\nu}$, and $T_{\mu\nu}$ is
the matter energy-momentum tensor. Considering a flat homogeneous
and isotropic FRW universe, we have
\begin{eqnarray} \nonumber \label{tetradasFRW}
e^i_\mu&=&{\rm diag}\left(1,a(t),a(t),a(t)\right), \\
e^\mu_i&=&{\rm
diag}\left(1,\frac{1}{a(t)},\frac{1}{a(t)},\frac{1}{a(t)}\right),
\end{eqnarray}
where $a(t)$ is the cosmological scale factor. By substituting Eqs.
(\ref{tetradasFRW}), (\ref{torsion2}), (\ref{S}) and (\ref{K}) into
Eq. (\ref{lagTele}), one could obtain the torsion scalar as
\cite{bengochea2009dark}
\begin{equation}  \label{STFRW}
T\equiv S^{\rho\mu\nu}T_{\rho\mu\nu}=-6H^2,
\end{equation}
where $H$ is the Hubble parameter $H=\dot{a}/a$. The dot represents
the first derivative with respect to the cosmic time. Substituting
Eq. (\ref{tetradasFRW}) into (\ref{ecsmovim}), one can obtain the
corresponding Friedmann equations
\begin{eqnarray}\label{f1}
12H^2 f_{T}+f=2k^2 \rho ,\\
\label{f2} 48 H^2 \dot{H}f_{TT}-(12H^2+4\dot{H})f_{T}-f=2k^2p,
\end{eqnarray}
where $\rho$ and $p$ are the total energy density and pressure,
respectively. By defining the effective energy density density
$\rho_{\rm eff}$, pressure $p_{\rm eff}$ and effective equation of
state (EoS) parameter $w_{\rm eff}$ as
\begin{eqnarray}
\label{rhoT}
\rho_{\rm eff} &=& \frac{1}{2k^2}(-12H^2f_T-f+6H^2), \\
\label{pT}
p_{\rm eff} &=& -\frac{1}{2k^2}[48\dot{H}H^2f_{TT}-4\dot{H}f_{T}+4\dot{H}] - \rho_{\rm eff}, \\
w_{\rm eff} &=& -
\frac{f/T-f_T+2Tf_{TT}}{\left[1+f_T+2Tf_{TT}\right]
\left[f/T-2f_T\right]}.
\end{eqnarray}
The Friedmann equations could be rewrite as
\begin{eqnarray}
\frac{3}{k^2}H^2 &=& \rho+\rho_{\rm eff}, \\
\frac{1}{k^2}(2\dot{H}+3H^2) &=& -(p + p_{\rm eff}),
\end{eqnarray}

Therefore, the cosmic acceleration could be driven by the torsion
instead of dark energy. In this cosmological framework, the
corresponding normalized Hubble parameter is
\begin{eqnarray}\label{Ez}
E^2(z)\equiv \frac{H^2(z)}{H^2_0}=\frac{T(z)}{T_0},
\end{eqnarray}
where $T_0=-6H^2_0$ (the subscript "0" denotes the current value).
Here we consider the matter and radiation in the Universe --- the
components whose energy density evolves with redshift $z$ as
$\rho_m=\rho_{m0}(1+z)^3$, $\rho_r=\rho_{r0}(1+z)^4$, respectively.
And then, Eq. (\ref{Ez}) could be expressed as
\cite{nunes2016new,nesseris2013viable}
\begin{equation}\label{Et}
E^2(z,\textbf{p})=\Omega_{m}(1+z)^3+\Omega_{r}(1+z)^4+\Omega_{F}y(z,\textbf{p})
\end{equation}
where $\Omega_{F}=1-\Omega_{m}-\Omega{r}$, and
$\Omega_{i}=\frac{k^2\rho_{i0}}{3H^2_0}$. In this way, a specific
form of $f(T)$ is embodied in the function $y(z,\textbf{p})$, whose
expression is
\begin{equation}\label{yz}
y(z,\textbf{p})=\frac{1}{T_0\Omega_{F}}\left(f-2Tf_T\right),
\end{equation}
where $\textbf{p}$ stands for the parameters in different forms of
$f(T)$ theory.

\subsection{Specific $f(T)$ models}

In this subsection we briefly review three specific $f(T)$ models,
which have passed basic observational tests \cite{nunes2016new}
and will be further investigated in this paper. \\

(1) The power-law model \cite{bengochea2009dark} (hereafter
$f_1$CDM) assumes that the Lagrangian density $f(T)$ of the theory
is the following:
\begin{equation}\label{ft1}
f(T)=\alpha(-T)^b
\end{equation}
where $\alpha$ and $b$ are two model parameters. The distortion
parameter $b$ quantifies deviation from the $\Lambda$CDM model,
whereas the parameter $\alpha$ can be expressed through the Hubble
constant and density parameter $\Omega_{F0}$ by inserting
Eq.~(\ref{ft1}) into Eq.~(\ref{Et}) with the boundary condition
$E(z=0)=1$ :
\begin{equation}
\alpha=(6H^2_0)^{1-b}\frac{\Omega_{F0}}{2b-1},
\end{equation}
Now Eq.~(\ref{yz}) may be rewritten as
\begin{equation}
y(z,b)=E^{2b}(z,b).
\end{equation}
Depending on the choice of parameter $b$, this $f(T)$ model can be
connected with some popular dark energy models. For $b=0$, it
 reduces to the $\Lambda$CDM, while it can mimic the
Dvali-Gabadadze-Porrati (DGP) model when $b=1/2$. \\

(2) The exponential model \cite{linder2010einstein} (hereafter
$f_2$CDM) is characterized by
\begin{equation}
f(T)=\alpha T_0(1-e^{-p\sqrt{T/T_0}}),
\end{equation}
where $\alpha$ and $p$ are two dimensionless parameters. Similarly
the expressions for $\alpha$ and $y(z,p)$ can also be obtained as
\begin{equation}
\alpha=\frac{\Omega_{F0}}{1-(1+p)e^{-p}},
\end{equation}

\begin{equation}\label{f2y}
y(z,p)=\frac{1-(1+pE)e^{-pE}}{1-(1+p)e^{-p}}.
\end{equation}
This model reduces to the $\Lambda$CDM in the limit $p\rightarrow +
\infty$. By setting $b=1/p$, Eq. (\ref{f2y}) is rewritten as
\begin{equation}
y(z,b)=\frac{1-\left(1+\frac{E}{b}\right)e^{-E/b}}{1-\left(1+\frac{1}{b}\right)e^{-1/b}}.
\end{equation}
and $\Lambda$CDM is recovered when $b\rightarrow 0^{+}$. \\

(3) Motivated by the exponential $f(R)$ gravity, the
hyperbolic-tangent model \cite{wu2011f} (hereafter $f_3$CDM) arises
from the ansatz
\begin{equation}
f(T)=\alpha(-T)^n\tanh\left(\frac{T_0}{T}\right)
\end{equation}
where $\alpha$ and $n$ are the two model parameters. We obtain the
expressions for $\alpha$ and $y(z,\textbf{p})$ as
\begin{eqnarray}
\alpha=-\frac{\Omega_{F0}(6H_{0})^{1-n}}{\left[ 2{\rm
sech}^{2}(1)+(1-2n){\rm tanh}(1)\right]},
\end{eqnarray}

\begin{eqnarray}
y(z,n)=E^{2(n-1)}\frac{2{\rm
sech}^{2}\left(\frac{1}{E^2}\right)+(1-2n)E^2{\rm
tanh}\left(\frac{1}{E^2}\right)}{2{\rm sech}^{2}(1)+(1-2n){\rm
tanh}(1)},
\end{eqnarray}
respectively. Compared with two previous $f(T)$ theories, this
$f(T)$ model cannot be reduced to the $\Lambda$CDM for any value of
its parameters. In addition, in order to have a positive value for
$\rho_{eff}$, the parameter $n$ must be greater than $3/2$
\cite{wu2011f}.

\section{Observational data and fitting method}

In order to measure the angular diameter distance, we always turn to
objects of known comoving size acting as ``standard rulers''. In
this paper, we will consider a combination of three types of
standard rulers using the most recent and significantly improved
observations, i.e., the compact radio quasars data from VLBI,
baryonic acoustic oscillations (BAO) from the large-scale structure,
and the cosmic microwave background (CMB) measurements.

\subsection{Quasars data}

It is well known that the baryon acoustic oscillations (BAO) peak
location is commonly recognized as a fixed comoving ruler of about
100 Mpc. Therefore it has already been used in cosmological studies
\cite{percival2010baryon,blake2011wigglez,beutler20116df}. In the
similar spirit, as extensively discussed in the literature, compact
radio sources (quasars, in particular) constitute another possible
class of standard rulers of about $10$ pc comoving length. Following
the analysis of Gurvits \cite{gurvits1994apparent}, luminosity and
redshift dependence of the linear sizes of quasars can be
parametrized as
\begin{equation}\label{lm}
l_m=lL^{\beta}(1+z)^n
\end{equation}
where $\beta$ and $n$ are two parameters quantifying the "angular
size - redshift" and "angular size - luminosity" relations,
respectively. The parameter $l$ is the linear size scaling factor
representing the apparent distribution of radio brightness within
the core. The data used in this paper were derived from an old 2.29
GHz VLBI survey undertaken by Preston et al.(1985), which contains
613 milliarcsecond ultra-compact radio sources covering the redshift
range $0.0035<z<3.787$. More recently, Cao et al.
\cite{cao2016cosmology} presented a method to identify a sub-sample
which can serve as a certain class of individual standard rulers in
the Universe. According to the optical counterparts and
luminosities, the full sample could be divided into three
sub-samples: low-luminosity quasars ($L<10^{27}$ W/Hz),
intermediate-luminosity quasars ($10^{27}$ W/Hz $<L<10^{28}$ W/Hz)
and high-luminosity quasars ($L>10^{28}$W/Hz). The final results
showed that only intermediate-luminosity quasars show negligible
dependence ($|n|\simeq 10^{-3}$, $\beta\simeq 10^{-4}$), and thus
they could be a population of rulers once the characteristic length
$l$ is fixed. In our analysis, we will use the observations of 120
intermediate-luminosity quasars covering the redshift range
$0.46<z<2.80$, while the linear size of this standard ruler is
calibrated to $l=11.03$ pc through a new cosmology-independent
technique \cite{cao2016cosmology}.

The observable quantity in this data-set is the angular size of the
compact structure in intermediate-luminosity radio quasars, whose
theoretical (i.e. determined by the cosmological model) counterpart
is
\begin{equation}
\theta_{th}(z)=\frac{l}{D_A(z)}
\end{equation}
where $D_A$ is the angular diameter distance at redshift $z$ and the
$f(T)$ model parameters $\textbf{p}$ directly enter the angular
diameter distance through
\begin{eqnarray}
\label{inted} D_A(z;\textbf{p})=\frac{3000h^{-1}}{(1+z)}\int_{0}^{z}
\frac{dz'}{E(z';\textbf{p})}
\end{eqnarray}
where $E(z';\textbf{p})$ is the dimensionless Hubble parameter and
$h$ is the dimensionless Hubble constant. We estimate the $f(T)$
parameters by minimizing the corresponding $\chi^2$ defined as
\begin{equation}
\chi^2_{QSO}(z;\textbf{p})=\sum_{i=1}^{120}\frac{[\theta_{th}(z_i;\textbf{p})-\theta_{obs}(z_i)]^2}{\sigma_{\theta}(z_i)^2}
\end{equation}
where $\theta_{obs}(z_i)$ is the observed value of the angular size
and $\sigma_{\theta}(z_i)$ is the corresponding uncertainty for the
$i$th data point in the sample. In order to properly account for the
intrinsic spread in linear sizes and systematics we have added in
quadrature $10\%$ uncertainties to the $\sigma_{\theta}(z_i)$.

\begin{table*}
\begin{center}
\begin{tabular}{|c| c| c| c |c|c|c|}
\hline
 $z_{BAO}$ &  $0.106$ &  $0.2$&  $0.35$& $0.44$& $0.6$& $0.73$ \\
\hline
 $\frac{r_s(z_d)}{D_V(z_{BAO})}$ & $ 0.336\pm0.015$ &  $0.1905\pm0.0061$& $ 0.1097\pm0.0036 $& $0.0916\pm0.0071$& $0.0726\pm0.0034$&  $0.0592\pm0.0032$\\
\hline
  $\frac{d_A(z_*)}{D_V(z_{BAO})}\; \frac{r_s(z_d)}{r_s(z_*)}$ &  $ 32.35\pm1.45$ &  $ 18.34\pm0.59$ &   $ 10.56\pm0.35$ &  $  8.82\pm0.68$ &  $ 6.99\pm0.33$ &   $ 5.70\pm0.31$\\
\hline
 $\frac{d_A(z_*)}{D_V(z_{BAO})}$ & $30.95\pm1.46$ &  $17.55\pm0.60$ & $10.11\pm0.37$ &  $ 8.44\pm0.67$&  $6.69\pm0.33$&  $5.45\pm0.31$\\
\hline
\end{tabular}
\caption{ Ratios of distances and the so called dilation scale
$D_V(z_{BAO})$ at different redshifts $z_{BAO}$ taken after
%\small{$\frac{r_s(z_d)}{D_V(z_{BAO})}$
\cite{percival2010baryon,blake2011wigglez,beutler20116df} and
\cite{giostri2012cosmic}.
%,$\frac{d_A(z_*)}{D_V(z_{BAO})}\;
%\frac{r_s(z_d)}{r_s(z_*)}$ and $\frac{d_A(z_*)}{D_V(z_{BAO})}$ for
%different values of $z_{BAO}$.}
\label{BAO}}
\end{center}
\end{table*}

\subsection{CMB and BAO data}

In order to diminish the degeneracy between $f(T)$ model parameters
we also used the accurate measurements of BAO and CMB.

The CMB experiments measure the temperature and polarization
anisotropy of the cosmic radiation in the early epoch. In general,
they are a very important tool for the inference of cosmological
model parameters. In particular, the shift parameter $R$ defined as:
\begin{eqnarray}
R=\sqrt{\Omega_{m}}\int_0^{z_*}\frac{dz'}{E(z';\textbf{p})},
\end{eqnarray}
where $z_*=1090.43$ denotes the decoupling redshift, is a convenient
quantity for a quick fitting of cosmological model parameters. The
first-year data release of \textit{Planck} reported its value of
$R=1.7499 \pm 0.0088$ \cite{ade2014planck}. We estimate the model
parameters by minimizing the corresponding $\chi^2$
\begin{eqnarray}
\chi_{CMB}^2=\left(\frac{R-1.7499}{0.0088}\right)^2.
\end{eqnarray}

The measurements of Baryon acoustic oscillation (BAO) in the
large-scale structure power spectrum and CMB angular power spectrum
have also been widely used for cosmological applications. In this
work we consider the measurements of $\frac{d_A(z*)}{D_V(z_{BAO})}$,
where $z*$ is the decoupling time,
$d_A(z)=\int^z_0\frac{dz'}{H(z')}$ is the co-moving angular-diameter
distance, and the dilation scale is given by
\begin{equation}
D_V(z)=\left(d_A(z)^2\frac{z}{H(z)}\right)^{1/3},
\end{equation}
The  BAO data are shown in Table \ref{BAO}. Similarly, the
corresponding $\chi^2$ for the BAO  probes is defined as
\begin{equation}
\chi_{BAO}^2=\sum_{ij} X_iC_{ij}^{-1}X_j,
\end{equation}
where
$X=\frac{d_A^{th}(z*)}{D_V^{th}(z_{BAO})}-\frac{d_A^{obs}(z*)}{D_V^{obs}(z_{BAO})}$
and $C_{ij}^{-1}$ is the inverse covariance matrix given by Ref.
\cite{giostri2012cosmic}.

\section{Observational constraints}

In this section, we determine the model parameters of three $f(T)$
cosmologies through the maximum likelihood method based on $\chi^2$
introduced in previous section using the Markov Chain Monte Carlo
(MCMC) method. Our code is based on CosmoMC
\cite{lewis2002cosmological} and we generated eight chains after
setting $R-1 = 0.001$ to guarantee the accuracy of this work.

\begin{figure*}
\centering
\includegraphics[width=8cm,height=6cm]{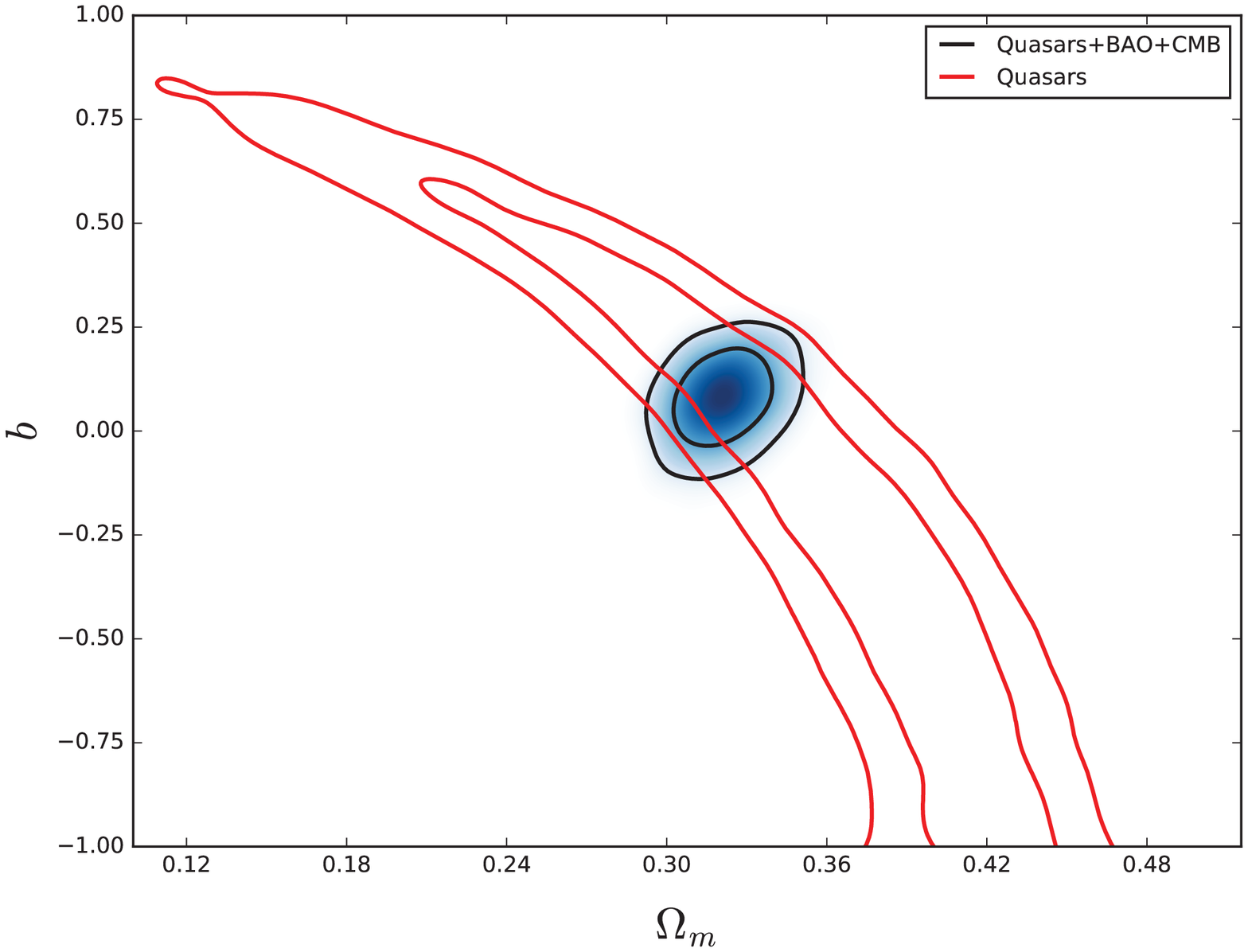}
\includegraphics[width=8cm,height=6cm]{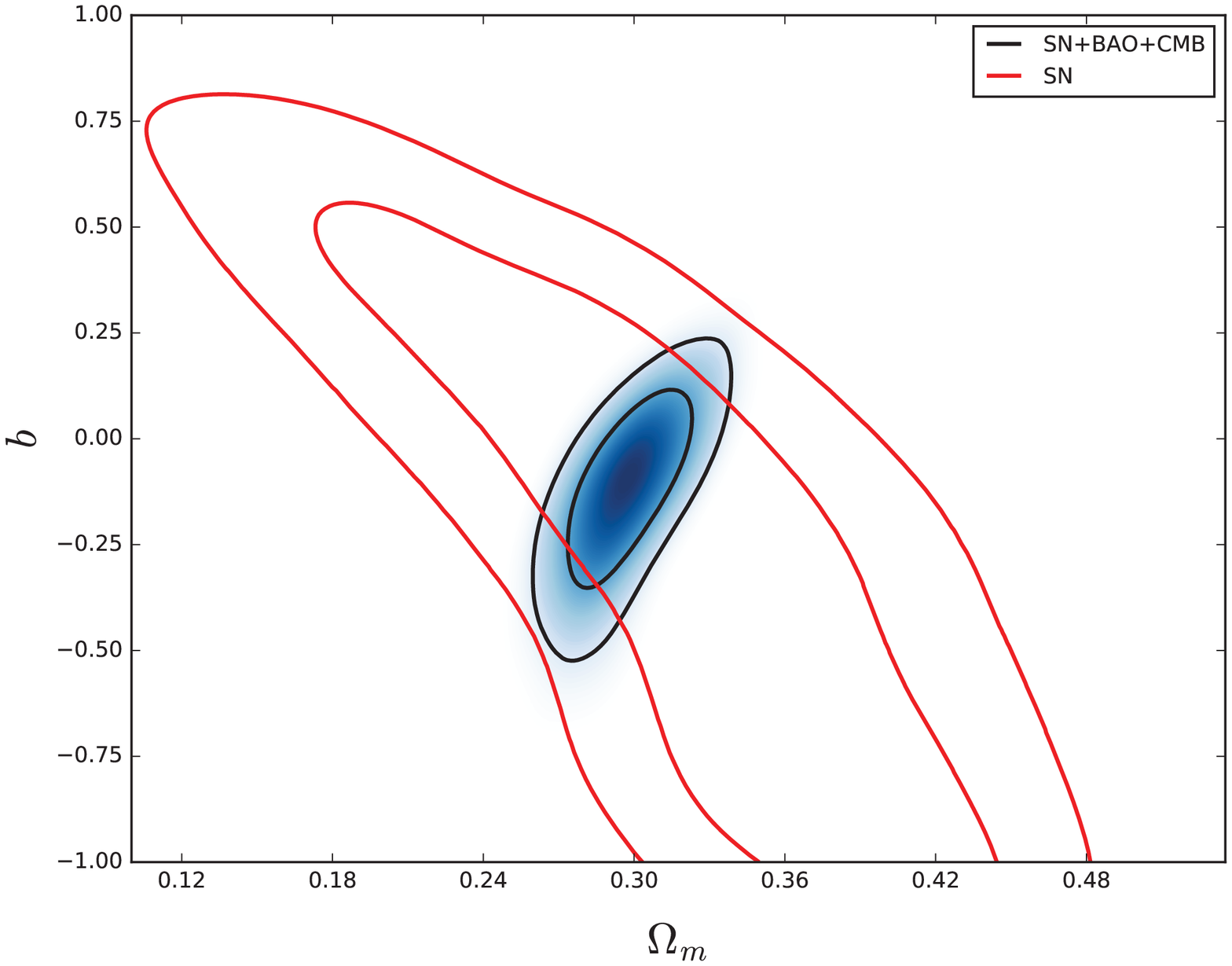}
\caption{$1 \sigma$ and $2 \sigma$ confidence regions for the
$f_1$CDM model. The red lines represent contour plot given by
quasars (left panel) and SN Ia (right panel). The black lines
represent constrained result from the joint analysis of
quasars+BAO+CMB (left panel) and SN Ia+BAO+CMB (right panel).
}\label{f1qs}
\end{figure*}

\begin{figure*}
\centering
\includegraphics[width=8cm,height=8cm]{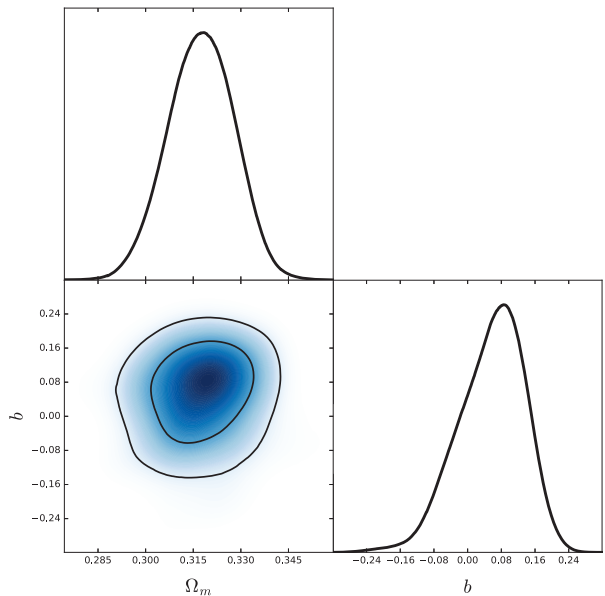}
\caption{The 68\% and 95\% confidence regions for the $f_1$CDM
model, which are constrained by the combined observational data of
quasars, SN Ia, BAO and CMB.}\label{1all}
\end{figure*}

\subsection{$f_1$CDM model: $f(T)=\alpha(-T)^b$}

In the case of the $f(T)$ theory based on $f(T)=\alpha(-T)^b$,
different data sets and their combinations led to the marginalized
2D confidence contours presented in Fig. \ref{f1qs}-\ref{1all}. The
corresponding marginal $1\sigma$ error bars can also be seen in
Table \ref{t1ft}.

Left panel of the Fig.~\ref{f1qs} shows the contours obtained from
the quasars only and in combination with CMB and BAO. We
remark that the quasar data only can not tightly constrain the model
parameters. In order to clearly illustrate the constraint comparison
between different data sets, a prior $b>-1$ is applied to the
likelihood contours obtained from the quasar data. Quantitatively,
the value of the distortion parameter $b$, which quantifies the
deviation from the $\Lambda$CDM model varies over the interval [-3,
0.56] within $1\sigma$ confidence level. As it is well known, the
main evidence for cosmic acceleration came from the other type of
distance indicators in cosmology, those probing the luminosity
distance, $D_L$ by observing the flux of type Ia supernovae (SN Ia).
In order to compare our fits with the results obtained using SN Ia,
likelihood contours obtained with the latest Union2.1 compilation
\cite{suzuki2012hubble} consisting of 580 SN Ia data points are also
plotted in the right panel of the Fig. \ref{f1qs}. It is clear that
the quasar data could give more stringent constraints than SN Ia,
and its constraining power becomes obvious when the large size
difference between the samples is taken into consideration. This may
happen due to the wider redshift range of the quasars data ($0.46< z
< 2.8$) compared with SN Ia ($0.015\leq z \leq 1.41$). Moreover, one
can clearly see from Fig. \ref{f1qs} that principal axes of
confidence regions obtained with SN and quasars are inclined at
higher angles, which sustains the hope that careful choice of the
quasar sample would eventually provide a complementary probe
breaking the degeneracy in the $f(T)$ model parameters. Finally, our
method based on the observations of intermediate-luminosity quasars
may also contribute to testing the consistency between luminosity
and angular diameter distances
\cite{cao2011a,cao2011b,cao2014cosmic}.

\begin{figure*}
\centering
\includegraphics[width=8cm,height=6cm]{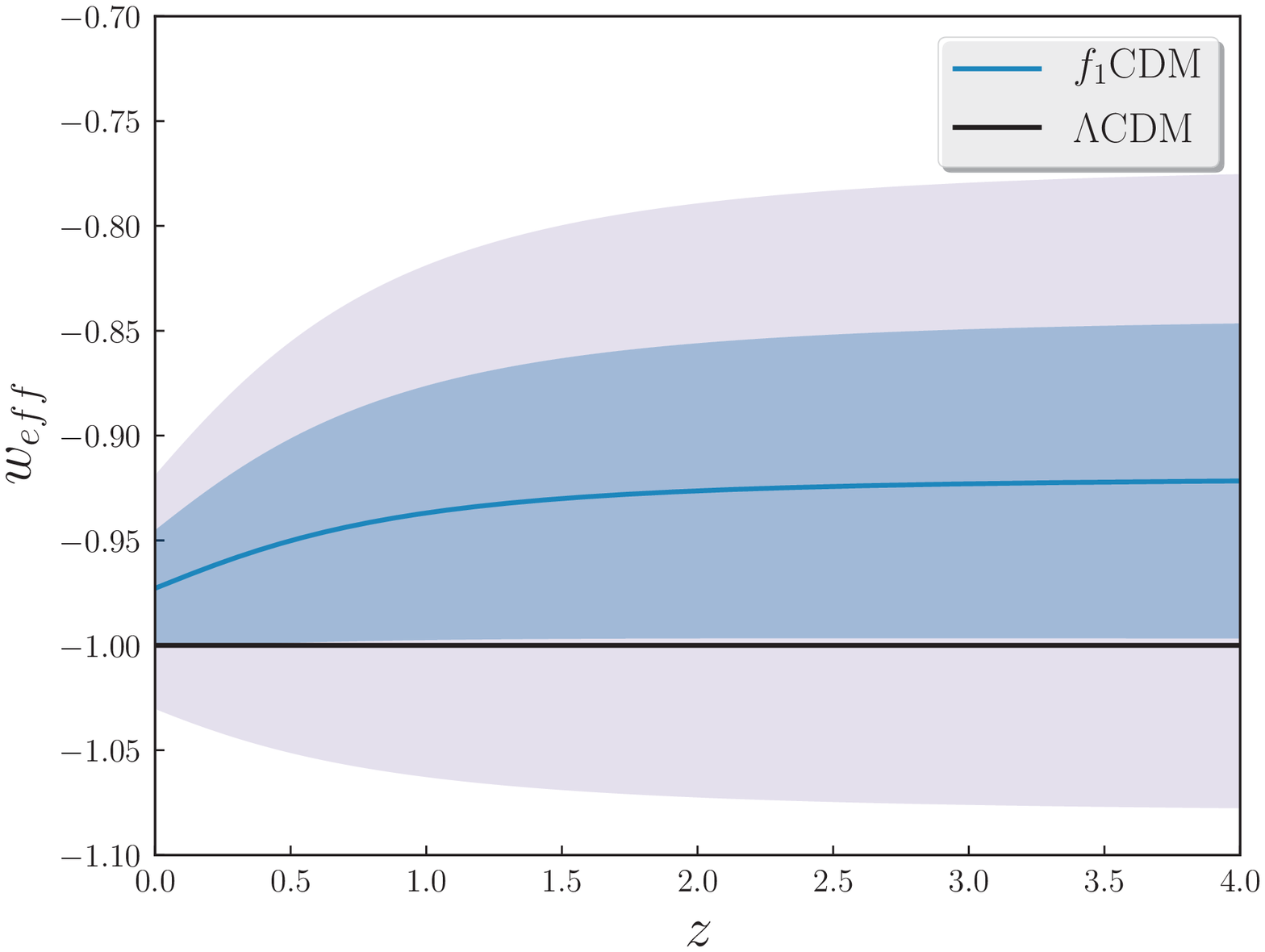}
\caption{ Evolution of the EoS for $\Lambda$CDM (black line)
and the effective EoS for the $f_1$CDM model (blue line) from the
joint analysis Quasars+BAO+CMB. $1\sigma$ and $2\sigma$
uncertainties are respectively denoted by blue and gray shades.
}\label{f1w}
\end{figure*}

With the combined standard ruler data sets of quasars, BAO and CMB,
the best-fit value for the parameters are $\Omega_m=0.321\pm 0.012$
and $b=0.080\pm 0.077$ within 68.3\% confidence level. For
comparison, fitting results from SN+BAO+CMB are also given in
Fig.~1. The best-fit value is $\Omega_m=0.297^{+0.015}_{-0.017}$ and
$b=-0.12^{+0.17}_{-0.13}$, which is in good agreement with that of
the Quasar+BAO+CMB data. It is obvious that the quasar data,
when combined to CMB and BAO observations, can give more stringent
constraints on this $f(T)$ cosmology, which demonstrates the strong
constraining power of BAO and CMB on the cosmological parameters.
This situation has also been extensively discussed in the previous
works investigating dark energy scenarios with other astrophysical
observations
\citep{cao2010testing,cao2011interaction,cao2011constraints,pan2012testing,cao2014cosmic}.
Again, the constraining power of 120 quasar data is comparable to
that of 580 SN Ia. On the one hand, the present value of the matter
density parameter $\Omega_m$ given by quasars is much lager than
that derived from other observations. This has been noted by our
previous analysis Cao et al. \cite{cao2016cosmology} and the
first-year \textit{Planck} results, in the framework of $\Lambda$CDM
cosmology. Such a result indicates that quasars data at high
redshifts may provide us a different understanding of the parameters
describing the components of the Universe. On the other hand, the
parameter $b$, which captures the deviation of $f(T)$ cosmology from
the $\Lambda$CDM scenario, seems to be vanishing or slightly larger
than 0 with the combined Quasar+BAO+CMB data. It is interesting to
note that $\Lambda$CDM is not included at $1\sigma$ confidence level
($b=0.08\pm 0.077$), this slight deviation from $\Lambda$CDM is also
consistent with a similar conclusion obtained in Ref.
\cite{nunes2016new} for this $f_1$CDM model. This tendency
can be more clearly seen from Fig.~\ref{f1w}, which illustrates the
comparison between the effective equation of state for $f(T)$ and
the EoS for $\Lambda$CDM model at $z\sim 4$, with the best-fitted
value as well as the 1$\sigma$ and 2$\sigma$ uncertainties derived
from the joint data of Quasars, BAO and CMB.

The contours constrained with the total combination of Quasars+SN
Ia+BAO+CMB are presented in Fig.~\ref{1all}, and the best-fit value
is $\Omega_m=0.317\pm 0.010$ and $b=0.057^{+0.091}_{-0.065}$. The
combined data give no stronger constraint, which indicates the
constraint ability of quasars data is already very strong, while SN
Ia do not play a leading role in the joint constraint. From the
results above, we can see the $\Lambda$CDM model which corresponds
to ($b=0$) is still included within 1$\sigma$ range. For comparison,
in Table \ref{t1ft} we also list alternative constraints obtained by
the others using different probes.

\begin{table*}[htb]
\centering
\begin{tabular}{ |l| c| c| l |}
\hline
Data & {\boldmath$\Omega_m$} & {\boldmath$b$} & Ref.\\
\hline
Quasars+BAO+CMB & $0.321\pm 0.012$ & $0.080\pm 0.077$ & This paper \\
\hline
SN Ia+BAO+CMB & $0.297^{+0.015}_{-0.017}$ & $-0.12^{+0.17}_{-0.13}$ & This paper \\
\hline
Quasars+SN Ia+BAO+CMB & $0.317\pm 0.010$ & $0.057^{+0.091}_{-0.065}$ & This paper\\
\hline\hline

OHD+SN Ia+BAO+CMB & $0.2335^{+0.016}_{-0.019}$ & $0.05128^{+0.025}_{-0.019}$ & \cite{nunes2016new} \\
\hline
SN Ia+BAO+CMB+dynamical growth data & $0.272\pm 0.008 $ & $-0.017\pm 0.083$ & \cite{nesseris2013viable} \\
\hline
SN Ia+BAO+varying fundamental constants &$0.294\pm 0.022$& $-0.119\pm 0.185$ & \cite{nunes2016observational} \\
\hline

\end{tabular}
\caption{Summary of the best-fit values of parameters for the
$f_1$CDM model with 1$\sigma$ uncertainties for different
observations (OHD is the abbreviation of the observational $H(z)$
data). \label{t1ft}}
\end{table*}

\begin{figure*}
\centering
\includegraphics[width=8cm,height=6cm]{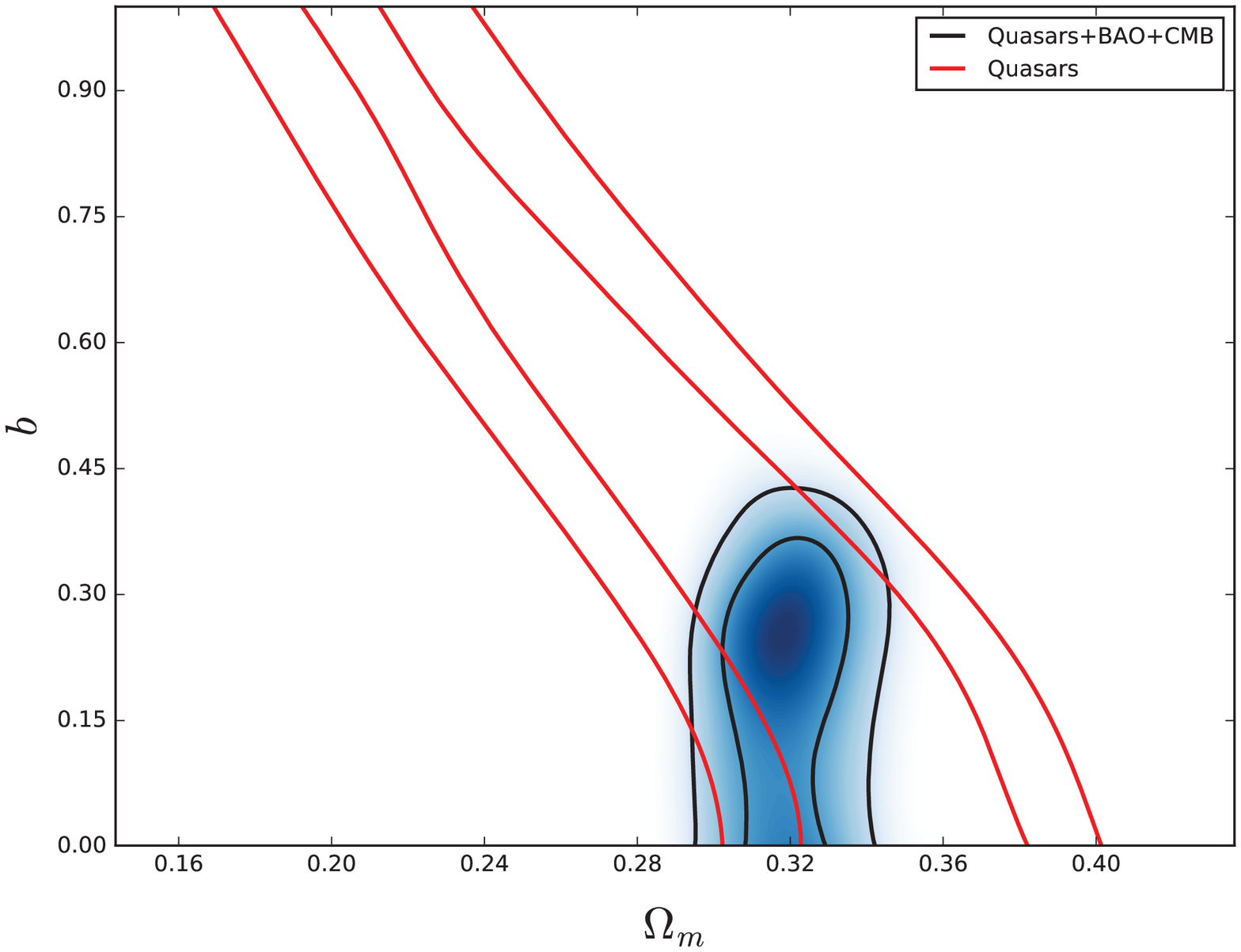}
\includegraphics[width=8cm,height=6cm]{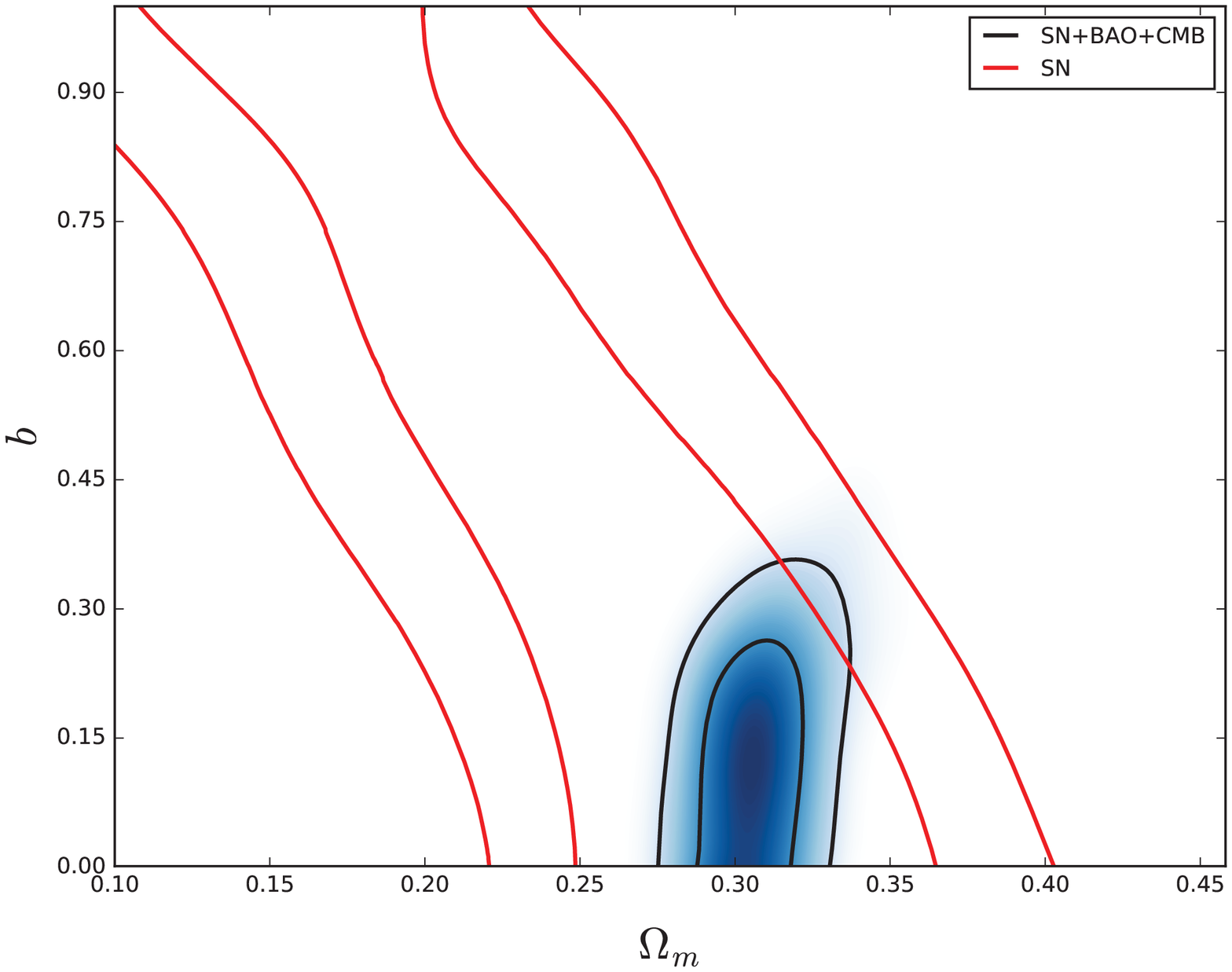}
\caption{$1\sigma$ and $2\sigma$ confidence regions for the $f_2$CDM
model. The red lines represent contour plot given by quasars (left
panel) and SN Ia (right panel). The black lines represent
constrained result from the joint analysis Quasars+BAO+CMB (left
panel) and SN Ia+BAO+CMB (right panel).}\label{f2qs}
\end{figure*}

\subsection{$f_2$CDM model: $f(T)=\alpha T_0(1-e^{-p\sqrt{T/T_0}})$}

Performing a similar analysis as before, this time with the other
$f(T)$ model in which $\Lambda$CDM is also nested, namely,
$f(T)=\alpha T_0(1-e^{-p\sqrt{T/T_0}})$, we made the same comparison
as $f_1$CDM discussed above, i.e. Quasars vs. SN Ia and
Quasars+BAO+CMB vs. SN Ia+BAO+CMB. The results are presented in
Fig.~\ref{f2qs} and the estimated cosmic parameters are briefly
summarized in Table \ref{t2ft}. It is apparent that the quasars data
exhibit similar constraining power as in the case of $f_1$CDM model,
which implies that the constraint ability of 120 quasar data can be
comparable to that of 580 SN Ia. By fitting the $f_2$CDM model to
Quasars+BAO+CMB, we obtain $\Omega_m = 0.319\pm 0.011$ and $b <
0.268$ (let us recall that here we introduced $b=1/p$).

\begin{figure*}
\centering
\includegraphics[width=8cm,height=6cm]{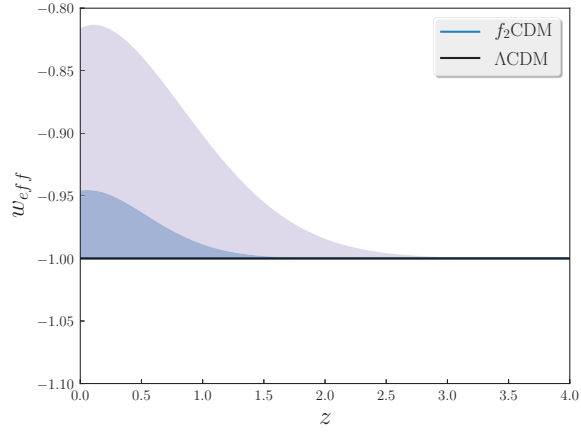}
\caption{ Evolution of the EoS for $\Lambda$CDM and the
effective EoS for the $f_2$CDM model from the joint analysis
Quasars+BAO+CMB. }\label{f2w}
\end{figure*}

\begin{figure*}
\centering
\includegraphics[width=8cm,height=8cm]{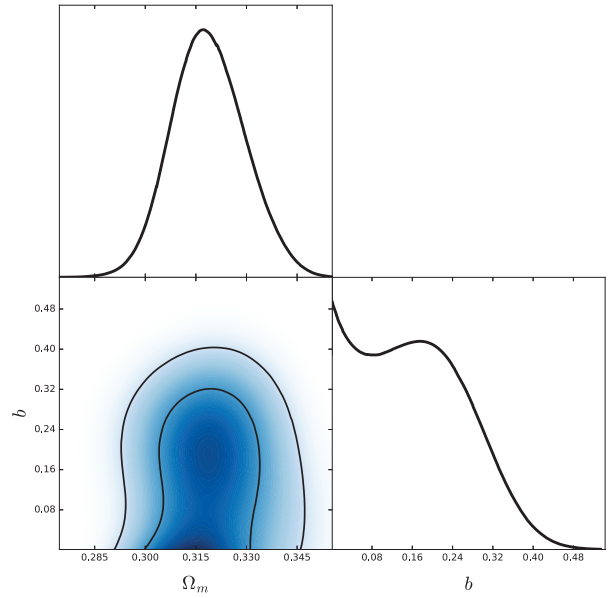}
\caption{The 68\% and 95\% confidence regions for the $f_2$CDM
model, which are constrained by the combined observational data of
Quasars, SN Ia, BAO and CMB.}\label{2all}
\end{figure*}

With the combined data set of Quasars+SN Ia+BAO+CMB, we also get the
marginalized 1$\sigma$ constraints of the parameters as
$\Omega_m=0.319^{+0.010}_{-0.011}$ and $b < 0.224$. The marginalized
1$\sigma$ and 2$\sigma$ contours of each parameter are presented in
Fig.~\ref{2all}. In Table \ref{t2ft}, the best-fit parameters and
their 1$\sigma$ uncertainties for three data sets are displayed. As
previously the results from the others using different probes are
shown for comparison. Obviously, the present matter density
parameter $\Omega_m$ fitted by quasars is lager than given by other
observations. The parameter $b$ quantifying the deviation
from the $\Lambda$CDM scenario, tends to be zero for all of
observations listed in Table \ref{t2ft}, which results in that the
exponential gravity is practically undistinguishable from
$\Lambda$CDM. As can be seen from the results presented in
Fig.~\ref{f2w}, even at 2$\sigma$ confidence level, the effective
EoS of $f_2$CDM model from the joint analysis of Quasars, BAO and
CMB agrees very well with that of $\Lambda$CDM at $z\sim 4$, which
strongly indicates the consistency between the two types of
cosmological models at much higher redshifts.

\begin{table*}
\centering
\begin{tabular}{ |l|c|c|l|}
\hline
Data & {\boldmath$\Omega_m$} & {\boldmath$b$} & Ref. \\
\hline

Quasars+BAO+CMB &  $0.319\pm 0.011$ & $b < 0.268$ & This paper \\
\hline
SN Ia+BAO+CMB & $0.307\pm 0.013$ & $b < 0.186$ & This paper\\
\hline
Quasars+SN Ia+BAO+CMB & $0.319^{+0.010}_{-0.011}$ & $b < 0.224$ & This paper \\
\hline \hline
OHD+SN Ia+BAO+CMB & $0.2784^{+0.0097}_{-0.019}$ & $0.1325^{+0.043}_{-0.13}$ & \cite{nunes2016new} \\
\hline
SN Ia+BAO+CMB+dynamical growth data & $0.272\pm 0.004 $ & $0.121\pm 0.184$ & \cite{nesseris2013viable} \\
\hline
SN Ia+BAO+varying fundamental constants &$0.283\pm 0.018$& $0.024\pm 0.08$ & \cite{nunes2016observational} \\
\hline
\end{tabular}
\caption{Summary of the best-fit values of parameters for the
$f_2$CDM model with 1$\sigma$ uncertainties for different
observations. \label{t2ft}}
\end{table*}

\subsection{$f_3$CDM model: $f(T)=\alpha(-T)^n\tanh\left(\frac{T_0}{T}\right)$}

Now we will discuss the third $f(T)$ cosmology which is truly an
alternative to the $\Lambda$CDM since the concordance cosmological
model cannot be recovered as a limiting case of $f_3$CDM model.
Consequently, the parameter $n$ does not characterize the deviation
from $\Lambda$CDM.

\begin{figure*}
\centering
\includegraphics[width=8cm,height=6cm]{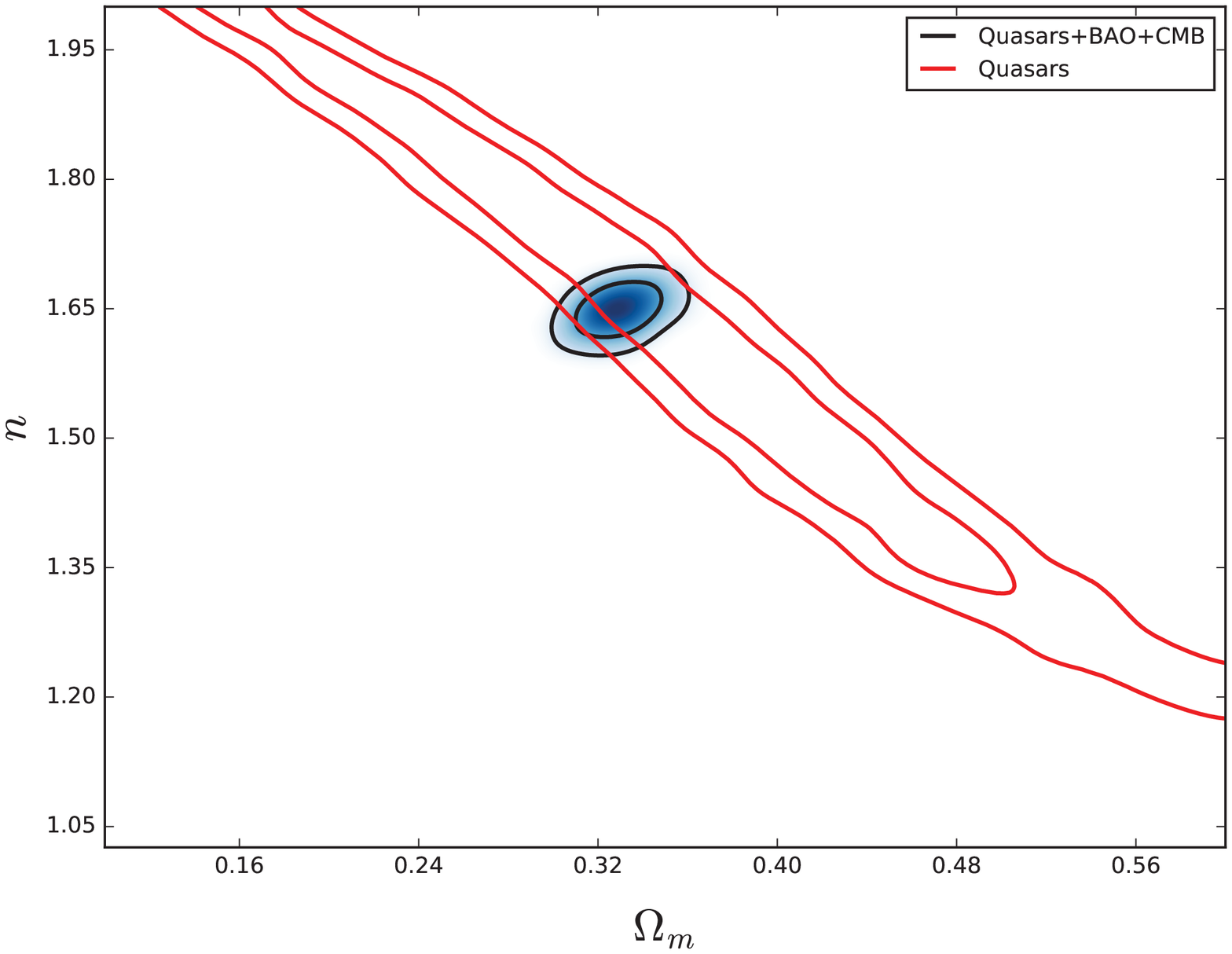}
\includegraphics[width=8cm,height=6cm]{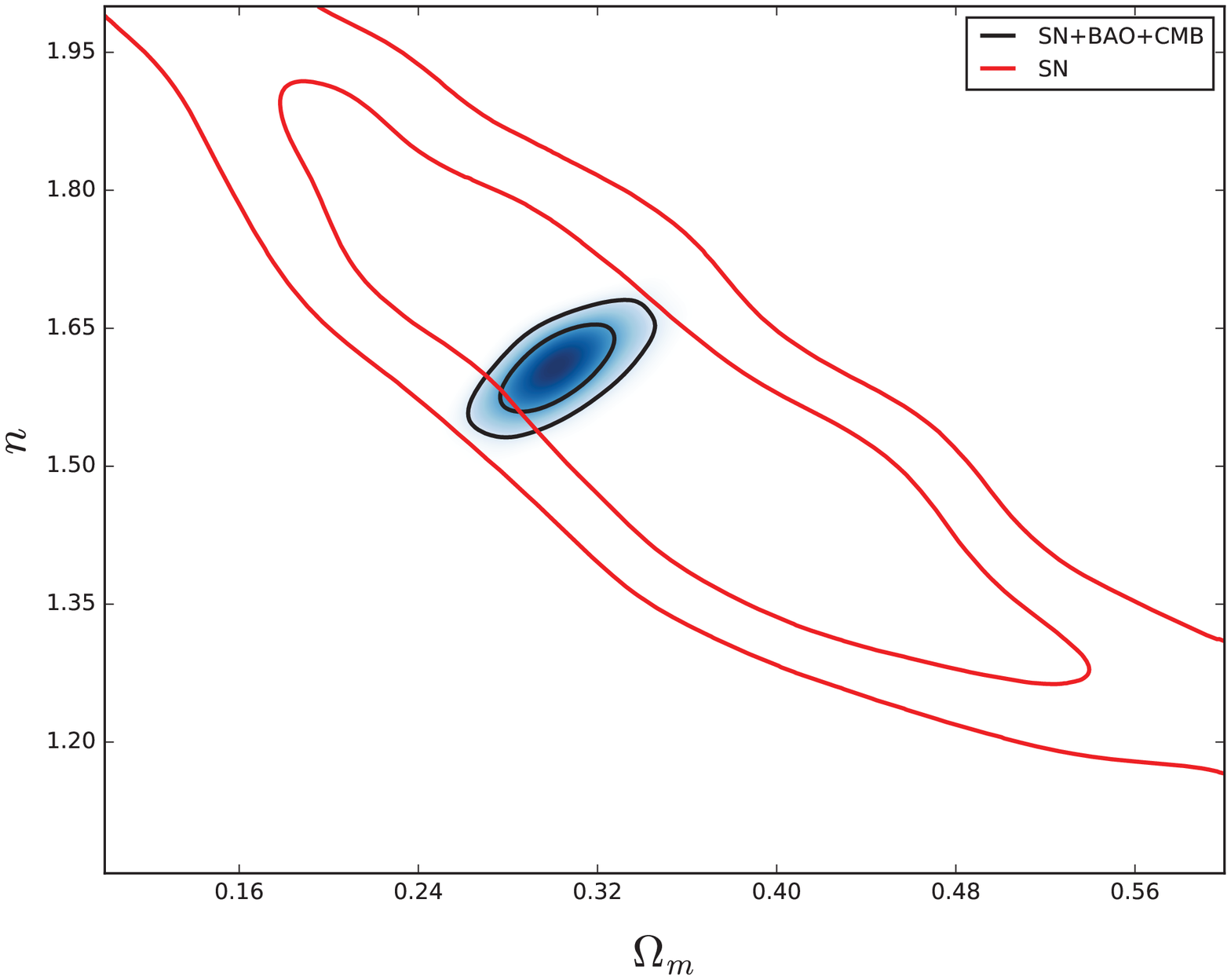}
\caption{$1 \sigma$ and $2 \sigma$ confidence regions for the
$f_3$CDM model. The red lines represent contour plot given by
Quasars (left panel) and SN Ia (right panel). The black lines
represent constrained result from the joint analysis Quasars+BAO+CMB
(left panel) and SN Ia+BAO+CMB (right panel).}\label{f3qs}
\end{figure*}

\begin{figure*}
\centering
\includegraphics[width=8cm,height=8cm]{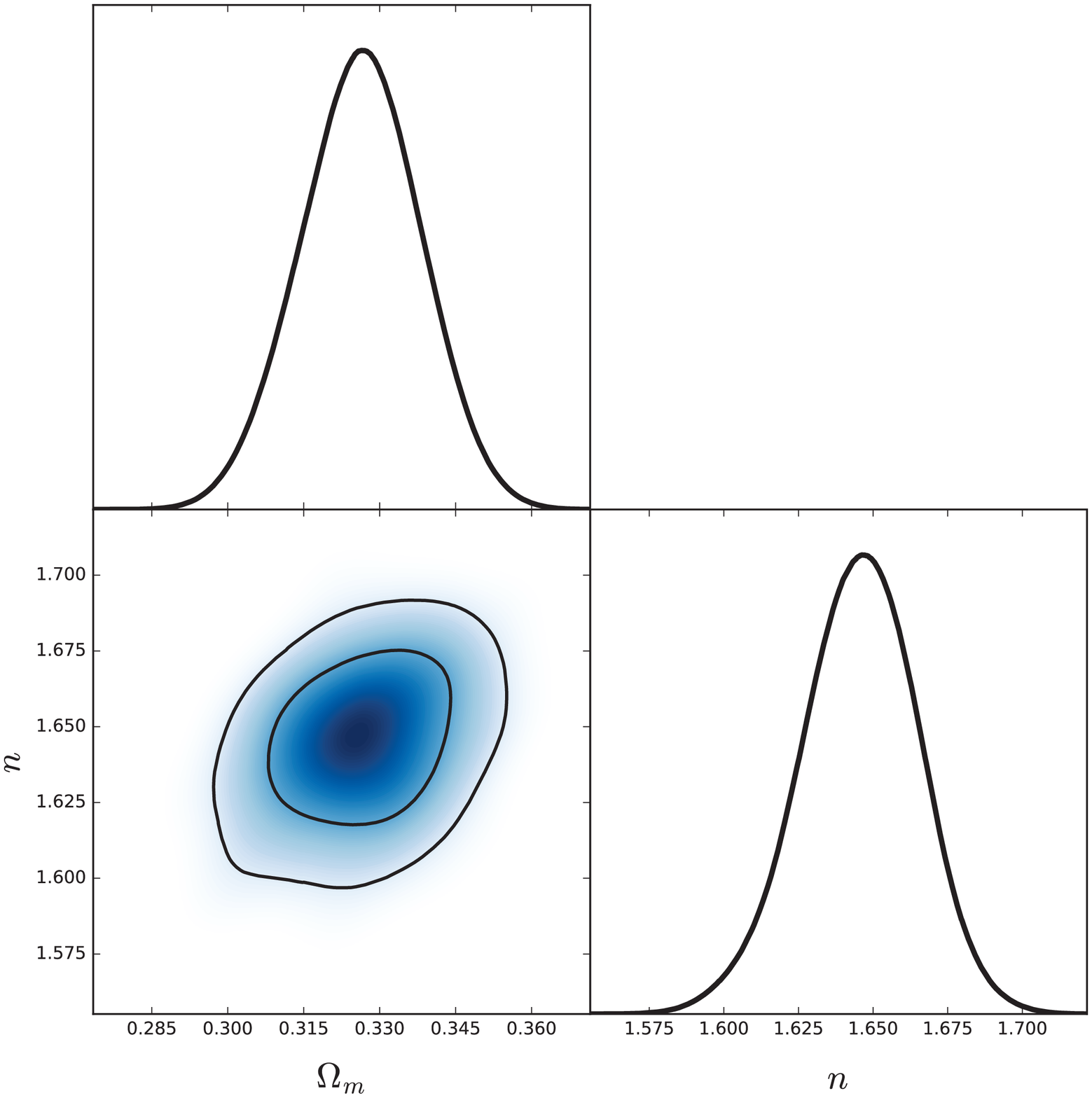}
\caption{The 68\% and 95\% confidence regions for the $f_3$CDM
model, which are constrained by the combined observational data of
Quasars, SN Ia, BAO and CMB.}\label{3all}
\end{figure*}

In Fig.~\ref{f3qs} we presented contour plots of $f_3$CDM model
parameters fitted to four different probes, namely Quasars, SN Ia,
Quasars+BAO+CMB, and SN Ia+BAO+CMB. As we can see, the quasar data
provide more stringent constraints than SN Ia, which indicates that
the constraining ability of quasar data can be comparable to or
better than that of SN Ia at least in this particular model. In
Fig.~\ref{3all} we show the contour plots for the combination of all
data sets Quasars+SN Ia+BAO+CMB. Additionally, in Table~\ref{t3ft}
we summarize the best-fit values for the three combined data sets
respectively. The table \ref{t3ft} also includes the best-fit values
and their 68\% confidence levels for the previous results from the
literature. Similar to the cases of $f_1$CDM model and $f_2$CDM
model, the present matter density parameter $\Omega_m$ implied by
quasars is lager than that given by other observations. Concerning
the value of the parameter $n$, its the value constrained by all of
the current observations satisfies the condition $n>3/2$, which is
necessary to achieve the cosmic acceleration in the framework of
$f_3$CDM.

\begin{figure*}
\centering
\includegraphics[width=8cm,height=6cm]{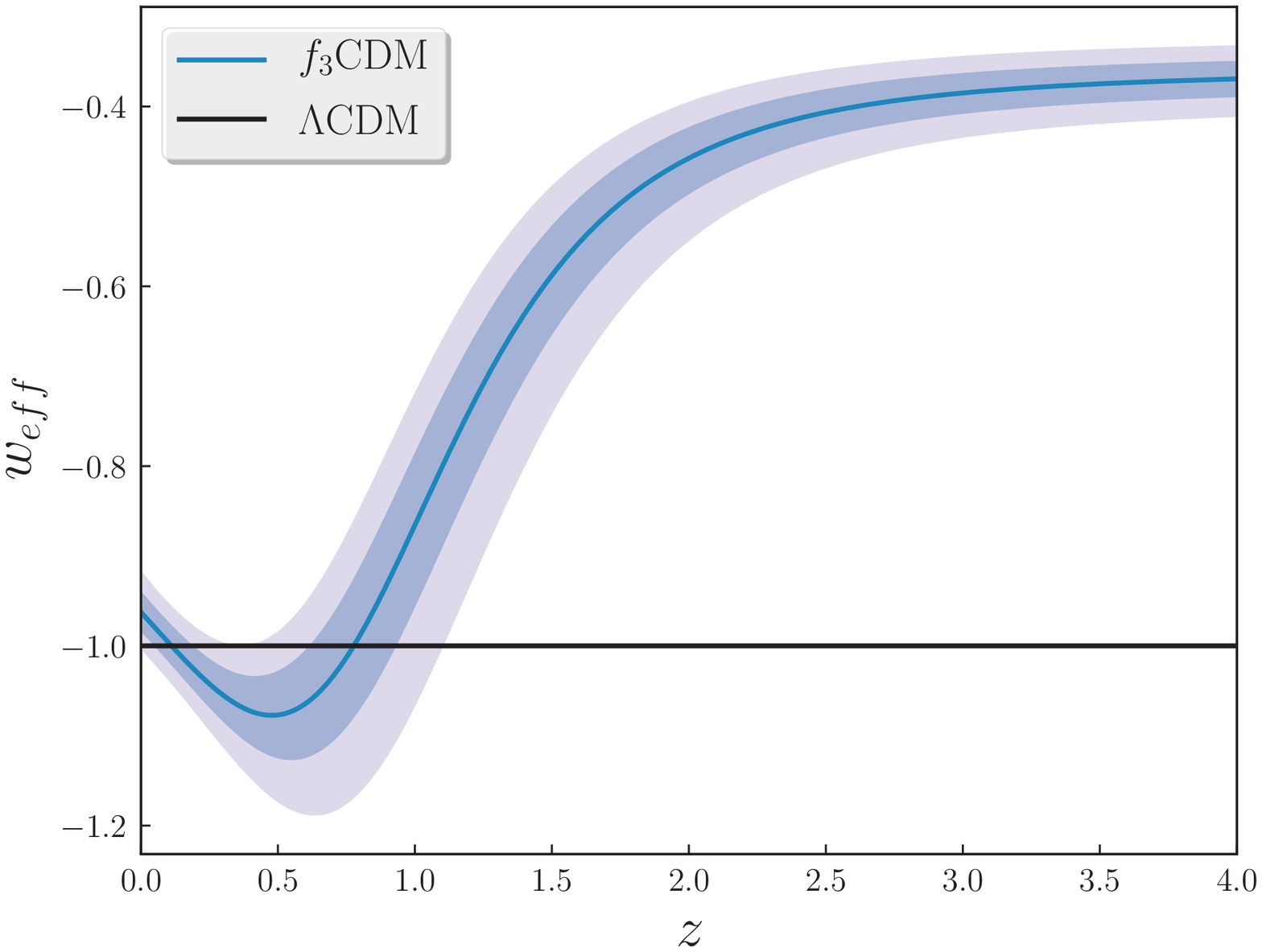}
\caption{ Evolution of the EoS for $\Lambda$CDM and the effective
EoS for the $f_3$CDM model from the joint analysis
Quasars+BAO+CMB.}\label{weff}
\end{figure*}

In Fig.~\ref{weff} we show the evolution of the effective equation
of state for $f_3$CDM model as a function of redshift, concerning
the best-fitted value with the 1$\sigma$ and 2$\sigma$ uncertainties
from the joint data of Quasars, BAO and CMB. In particular, we find
that the value of $n$ obtained with quasars suggests that the
effective equation of state crosses the phantom divide line at lower
redshifts \cite{wu2011f}.

\begin{table*}[htb]
\centering
\begin{tabular}{ |l|c|c|l|}
\hline
Data & {\boldmath$\Omega_m$} & {\boldmath$n$} & Ref. \\
\hline
Quasars+BAO+CMB & $0.329\pm 0.011$ & $1.649\pm 0.021$ & This paper \\
\hline
SN Ia+BAO+CMB & $0.303\pm 0.017$ & $1.607\pm 0.031$ & This paper \\
\hline
Quasars+SN Ia+BAO+CMB & $0.326\pm 0.012$ & $1.645^{+0.020}_{-0.018}$ & This paper \\
\hline\hline
GRB+OHD+SN Ia+BAO+CMB & $0.286^{+0.013}_{-0.012}$ & $1.616^{+0.02}_{-0.035}$ & \cite{cardone2012accelerating} \\
\hline
\end{tabular}
\caption{Summary of the best-fit values of parameters for the
$f_3$CDM model with 1$\sigma$ uncertainties for different
observations.\label{t3ft}}
\end{table*}

\subsection{Model selection}

In order to to make a good comparison between different
models or decide which model is preferred by the observational data,
we will use two standard information criteria, namely the Akaike
Information Criterion (AIC) \citep{akaike1974a} and the Bayesian
Information Criterion (BIC) \citep{schwarz1978estimating} to study
competing models. The above two information criteria are
respectively defined as
\begin{equation}
\rm{AIC}=-2\ln\mathcal{L}+2k,
\end{equation}
and
\begin{equation}
\rm{BIC}=-2\ln\mathcal{L}+k\ln{N},
\end{equation}
where $\mathcal{L}=\exp(-\chi^2_{min}/2)$, $k$ represents the number
of free parameters in the model and $N$ is the sample size used in
the statistical analysis. In addition, we introduce the ratio of
$\chi^2_{min}$ to the degrees of freedom (d.o.f),
$\chi^2_{min}/d.o.f$, to judge the quality of observational data
set.

In Table~\ref{aic}, we list the values of AIC, BIC and
$\chi^2_{min}/d.o.f$ for different models from the joint analysis
Quasar+BAO+CMB and SN Ia+BAO+CMB. It is obvious that both of AIC and
BIC criteria support $\Lambda$CDM to be the best cosmological model
consistent with the available observations, since the IC value it
yields is the smallest. Concerning the ranking of the three $f(T)$
models, AIC and BIC criteria tend to provide the same conclusions as
follows. The $f_2$CDM model performs the best in explaining the
current data, which can be clearly seen from the similarity between
$f_2$CDM and $\Lambda$CDM shown in Fig.~\ref{f2w}. Then next after
$f_2$CDM is the $f_1$CDM model, which can also reduce to the
$\Lambda$CDM model and its best-fit parameters indeed do so. The
worst model according to the AIC and BIC criteria is $f_3$CDM, which
is unable to provide a good fit to the data and can not nest
$\Lambda$CDM.

\begin{table*}[htb]
\centering
\begin{tabular}{ |l|c|c|c|c|c||c|c|c|c|c|}
\hline
&\multicolumn{5}{|c||}{Quasar+BAO+CMB}&\multicolumn{5}{|c|}{SN Ia+BAO+CMB} \\
\hline
 Model & AIC & $\Delta$AIC & BIC & $\Delta$BIC &$\chi^2_{min}/d.o.f$ & AIC & $\Delta$AIC & BIC & $\Delta$BIC&$\chi^2_{min}/d.o.f$\\
\hline
$\Lambda$CDM & $613.78$ & $0$ & $616.62$ & $0$& $4.80$ &$550.87$&$0$&$555.24$ & $0$& $0.95$\\
\hline
$f_1$CDM & $615.46$ & $1.68$ & $621.15$ & $4.53$& $4.81$ &$552.94$& $2.07$& $561.69$ & $6.45$& $0.95$ \\
\hline
$f_2$CDM & $615.32$ & $1.54$ & $621.01$ & $4.39$& $4.81$ &$552.83$ &$ 1.96$ & $561.58$ & $6.34$& $0.95$ \\
\hline
$f_3$CDM & $616.91$ & $3.13$ & $622.60$ & $5.98$& $4.83$ &$553.01$ &$2.14$ & $561.76$ &$6.52$& $0.95$\\
\hline
\end{tabular}
\caption{ Summary of the AIC and BIC values for different
models obtained from the combined Quasar+BAO+CMB data and the
combined SN Ia +BAO+CMB data.\label{aic}}
\end{table*}

\section{Conclusions and discussions}

As an interesting approach to modify gravity, $f(T)$ theory based on
the concept of teleparallel gravity, was proposed to explain the
accelerated expansion of the Universe without the need of dark
energy. In this paper, we have used the recently-released sample of
VLBI observations of the compact structure in 120
intermediate-luminosity quasars ($0.46<z<2.80$) to get the
constraints on the viable and most popular $f(T)$ gravity models.
The statistical linear sizes of these quasars observed at 2.29 GHz
show negligible dependence on redshifts and intrinsic luminosity,
and thus represent a fixed comoving-length of the standard ruler.
Therefore, the other motivation of this work was to investigate the
constraining ability of quasar data in the context of $f(T)$ models.
In particular, we have considered three $f(T)$ models with two
parameters, out of which two could nest the concordance $\Lambda$CDM
model and we quantifed their deviation from $\Lambda$CDM cosmology
through a single parameter $b$. For the third $f(T)$ cosmology which
can not be directly reduced to $\Lambda$CDM, we discussed the
possibility for the effective equation of state to cross the phantom
divide line.

In our investigation we have used (i) the very recently released
"angular size - redshift" data sets of 120 intermediate-luminosity
quasars in the redshift range $0.46< z <2.76$, (ii) the cosmic
microwave background and baryon acoustic oscillation data points.
Meanwhile, in order to compare our fits obtained with 120 quasars
(standard rulers), to the similar constraints obtained with the
Union 2.1 compilation consisting of 580 SN Ia data points (standard
candles) we also carried out respective analysis based on SNIa data.
Here we summarize our main conclusions in more detail:

\begin{itemize}

\item For all of the three the $f(T)$ models, all of the fitting results
show that the quasar data (N=120) could provide more stringent constraints than the
Union2.1 SN Ia data (N=580). This may be associated with the wider
redshift range covered by the quasar data ($0.46< z < 2.8$) compared
with SN Ia ($0.015\leq z \leq 1.41$). The constraining power of the
former becomes obvious when the large size difference between the
samples is taken into consideration. Moreover, one can clearly see
that principal axes of confidence regions obtained with SN and
quasars are inclined at higher angles, which sustains the hope that
careful choice of the quasar sample would eventually provide a
complementary probe breaking the degeneracy in the $f(T)$ model
parameters. Our method based on the observations of
intermediate-luminosity quasars may also contribute to testing the
consistency between luminosity and angular diameter distances.

\item The present value of the matter density parameter
$\Omega_m$ implied by quasars is much lager than that derived from
other observations, which has been noted by our previous analysis
and the first-year \textit{Planck} results, in the framework of
$\Lambda$CDM cosmology. Such result indicates that quasar data at
high redshifts may provide us a different understanding of the
components in the Universe.
  \item For $f_1$CDM and $f_2$CDM models, deviation from $\Lambda$CDM cosmology
is also allowed in the obtained confidence level, although the
best-fit value is very close to its $\Lambda$CDM one. It is
interesting in the present work to note that $\Lambda$CDM is not
included at $1\sigma$ confidence level for the power-law model
$f_1$CDM model, this slight deviation from $\Lambda$CDM is also
consistent with a similar conclusion obtained in the previous
observational studies on $f(T)$ gravity. In the framework of
$f_3$CDM, the value of $n$ constrained by all of the current
observations satisfies the limit of $n>3/2$, which is necessary to
achieve the cosmic acceleration. Moreover, we find that the value of
$n$ obtained with quasars suggests that the effective equation of
state can cross the phantom divide line at lower redshifts .

\item The information criteria (AIC and BIC) demonstrate that, compared with other three $f(T)$ scenarios
considered in this paper, the cosmological constant model is still
the best cosmological model consistent with the available
observations. Concerning the ranking of the $f(T)$ cosmologies, the
$f_2$CDM model performs the best in explaining the current data,
while the $f_3$CDM model gets the smallest support and can not nest
the concordance $\Lambda$CDM model. 

\item In summary, using for the recently released quasar data acting as a new source of standard rulers,
we were able to set more stringent limits on the viable and most
used $f(T)$ gravity models. Our results highlight the importance of
quasar measurements to provide additional information of various
candidates for modified gravity, especially the possible deviation
from $\Lambda$CDM cosmology. More importantly, given the
usefulness of this angular size data in pinning down parameter
values, we also anticipate that near-future quasar observations will
provide significantly more restrictive constraints on other
torsional modified gravity theories
\citep{kofinas2014cosmological,harko2014nonminimal,boisseau2000reconstruction}.

\end{itemize}

\begin{acknowledgements}
This work was supported by the National Key
Research and Development Program of China under Grants No.
2017YFA0402603; the Ministry of Science and Technology National
Basic Science Program (Project 973) under Grants No. 2014CB845806;
the National Natural Science Foundation of China under Grants Nos.
11503001, 11373014, and 11690023; the Fundamental Research Funds for
the Central Universities and Scientific Research Foundation of
Beijing Normal University; China Postdoctoral Science Foundation
under grant No. 2015T80052; and the Opening Project of Key
Laboratory of Computational Astrophysics, National Astronomical
Observatories, Chinese Academy of Sciences. This research was also
partly supported by the Poland-China Scientific \& Technological
Cooperation Committee Project No. 35-4. M.B. was supported by
Foreign Talent Introducing Project and Special Fund Support of
Foreign Knowledge Introducing Project in China.
\end{acknowledgements}

\bibliographystyle{spphys}
\bibliography{ftmodel}

\end{document}